\documentstyle[prd,aps,psfig]{revtex}
\begin{document}
\draft
\title{QUANTUM CORRECTIONS TO MAXWELL ELECTRODYNAMICS IN A HOMOGENEOUS AND ISOTROPIC UNIVERSE
WITH COSMOLOGICAL CONSTANT}
\author{M. R. de Garcia Maia\thanks{Electronic address: mrgm@dfte.ufrn.br},
J. C. Carvalho\thanks{Electronic address: carvalho@dfte.ufrn.br},
and C. S. da C\^amara Neto\thanks{Electronic
address: calistrato@dfte.ufrn.br}}
\address{Departamento de F\'{\i}sica, Universidade Federal do Rio
Grande do Norte, 59072-970 Natal RN Brazil}
\maketitle
\begin{abstract}
Some cosmological consequences of first order quantum corrections
to Maxwell electrodynamics are investigated in the context of
a spatially flat homogeneous and isotropic universe driven by
a magnetic field plus a cosmological
term $\Lambda$.
The introduction of these quantum corrections may provide a more
realistic model for the universe evolution.
For a vanishing $\Lambda$,
we derive the general solution corresponding to the particular
one recently found by Novello {\em et al.} [gr-qc/9806076]. We
also find a general solution for the case when $\Lambda$ is a
non-vanishing constant. Both solutions describe a non-singular,
bouncing universe that begins arbitrarilly large, contracts to a minimum
non-zero size $a_{min}$ and expands thereafter. However, we show that the
first order correction to the electromagnetic Lagrangean density,
in which the analysis is based, fails to describe the dynamics near
$a_{min}$, since, at this point,  the magnetic fields grows beyond the maximum
strength allowed by the approximation used ($B\ll 8.6\times
10^{-7}\:{\rm Tesla}=0.0086\:{\rm Gauss}$).
The time range where the first order approximation can be used is
explicitly evaluated. These problems my be
circumvented through the use of higher order terms in the effective Lagrangean,
as numerical calculations performed by Novello {\em et
al.} [gr-qc/9809080], for the vanishing $\Lambda$ case, have
indicated.
They could also be evaded
in some models based on oscillatory behaviour of the fundamental constants.
A third general solution
corresponding to a constant magnetic field sustained by a time dependent
$\Lambda$ is derived. The temporal behaviour of $\Lambda$ is
univocally determined. This latter solution is
capable of describing the whole cosmic history and describes a
universe that, although with  vanishing curvature ($K=0$), has a
scale factor that approaches zero asymptotically in
the far past, reaches a maximum and then contracts back to an
arbitrarilly small size. The cosmological term decays during the
initial expansion phase and increases during the late contraction
phase, so as to keep $B$ constant throughout.
An important feature of this model is that it
presents an inflationary dynamics except in a very short period of time
near its point of maximum size.
\end{abstract}
\section{INTRODUCTION}
\label{s1}
In a recent paper, Novello and colaborators \cite{novello98a} have
analysed the cosmological consequences of quantum corrections to Maxwell
electrodynamics. They have considered the quantum effects leading to the production of
electron-positron pairs that were first derived by Heisenberg and Euler
\cite{heisenberg36}. The analysis was a semiclassical one (quantum field in a classical general
relativistic geometry), made in first order on the
effective Lagrangean density (weak field limit) and applied to  a spatially
flat Friedmann-Robertson-Walker (FRW) model.
The more interesting
result found in Ref. \cite{novello98a} was the
removal of the primordial singularity due to the appearance of a
negative pressure in the early stages of the universe. The
analytical
solution derived shows that the energy density
associated with the electromagnetic field vanishes at the point
where the scale factor reaches its minimum. The
non-singular behaviour of the model is unaffected by the presence
of other types of ultrarelativistic matter obeying the equation of
state $p_{(ur)} = \rho_{(ur)} /3$. In a subsequent paper \cite{novello98b}, the
analysis was extended beyond the first order approximation and
the numerical solutions obtained show that the non-singular
behaviour is preserved.

The conclusion reached by the authors of references
\cite{novello98a,novello98b} was that the cosmological singularity
of FRW models is a distinguished feature of classical
electrodynamics. This problem is overcome when quantum
corrections are taken into account, leading to a more realistic
description of the universe.

In the present paper we extend the analysis developed
in \cite{novello98a}. We begin by showing that the analytical
solution obtained in that paper is a particular one and write down
the corresponding general solution. We then analyse how this solution is modified by
the presence of a non-zero cosmological term $\Lambda$ in two different
situations: we consider both the case of a constant $\Lambda$,
which leads also to a non-singular universe,
and the case of a time varying ``cosmological constant'' that
supplies energy to a constant cosmological magnetic field. The
equation that leads to a constant magnetic field as a possible
solution to the field equations appeared already in Ref.
\cite{novello98a} but was disregarded. The presence of a
time varying $\Lambda$ gives physical meaning to such a solution.
The time dependence for $\Lambda$ is univocally determined and
represents a slightly modification of a form that has been
suggested frequently in the literature ($\Lambda = 3H^2 - c_o$, $c_0=$ constant
\cite{overduin98}).
The universe described
by this latter solution begins arbitrarily small in the far past,
expands towards a maximum size $a_{max}$ at the point where the vacuum
energy reaches its minimum and then contracts towards an
arbitrarily small size in the distant future. A remarkable feature
of this model is that it has an inflationary dynamics, except
during a short period of time near $a_{max}$. This scenario
obviously violate the
standard evolution law for the magnetic field ($B(t) \propto
a^{-2}$), which is commonly used to rescale to present time the several constraints
on primordial magnetic fields presented in the literature
\cite{suh98,olinto98}.

We also discuss the restrictions on the domain of validity of the
solutions found, due to the use of the
first order approximation for the effective electromagnetic Lagrangean.

The paper is organized as follows: In Section \ref{s2} we set down the basic
equations. In  Section \ref{s3} we generalize the
solution derived in \cite{novello98a} which assumes a vanishing cosmological term
and a time-dependent magnetic field.. In Section \ref{s4} we obtain a
new solution that takes into account the presence of a constant
$\Lambda$. A new solution involving the presence of a $\Lambda(t)$
plus a constant magnetic field is derived is Section \ref{s5}. Section
\ref{s6}
contains a summary of our results and suggestions for future work.

In what follows Greek indices run from 0 to 4 and Latin indices run from 1 to
3.
As in \cite{novello98a}, we use Heaviside-Lorentz  electromagnetic units, but,
unlike that reference, and unless otherwise stated, we make $\hbar =c = 1$, . This is the scheme
used in \cite{kT} which seems to be more appropriate to our needs.
In this system, the magnetic field $\vec{B} = \vec{H}$ \cite{jackson} is measured
in Tesla \cite{kT} and the fine structure constant is given  by \cite{kT}
\begin{equation}
\label{e1}
\alpha = \frac{e^2}{4\pi}=\frac{1}{137.036}\;,
\end{equation}
so that the magnitude of the electron charge is
\begin{equation}
\label{e2}
e=\sqrt{4\pi\alpha} = 0.30282\;.
\end{equation}
(See the Appendix A of Ref. \cite{kT} and the Appendix on Units
and Dimensions of Ref. \cite{jackson} for further details.)
\section{FUNDAMENTAL EQUATIONS}
\label{s2}
In the system of units used in this paper, the Lagrangean density for
free fields in classical Maxwell electrodynamics is written as
\begin{equation}
\label{e3}
\protect{\cal{L}}_{(MAXWELL)} =-\frac{1}{4}F^{\mu\nu}F_{\mu\nu} = -\frac{1}{4}F\;,
\end{equation}
where $F^{\mu\nu}$ is the electromagnetic field strength tensor.
Canonical energy-momentum tensor is then given by
\begin{equation}
\label{e4}
T_{\mu\nu}^{(MAXWELL)} =
F_{\mu\alpha}{F^{\alpha}}_{\nu}+\frac{1}{4}Fg_{\mu\nu}\;.
\end{equation}

The quantum corrections to classical electrodynamics that should
be considered in order to describe the process of creation of electron-positron pairs
by the electromagnetic field were evaluated by Heisenberg and Euler
\cite{heisenberg36,schwinger51}. A clear derivation of the
effective Lagrangean that describes these quantum effects can be
found in \cite{gMm}, where one should be aware that the Gaussian system of
electromagnetic units is used. The first order calculation yields
for the effective Lagrangean density \cite{novello98a,gMm}
\begin{equation}
\label{e5}
\protect{\cal{L}}=-\frac{1}{4}F +\mu \left [\frac{1}{4}F^2
+\frac{7}{16}(F^{*})^2\right ]\;,
\end{equation}
where
\begin{equation}
\label{e6}
F^{*}\equiv F^{*}_{\mu\nu}F^{\mu\nu}\;,
\end{equation}
$F^{*}_{\mu\nu}$ is the dual of $F_{\mu\nu}$,
\begin{equation}
\label{e7}
\mu\equiv \frac{8}{45}\frac{\alpha^2}{m_e^4}\approx 1.4\times 10^8 \,{\rm GeV}^{-4}\;,
\end{equation}
and $m_e = 5.110 \times 10^{-4}\; {\rm GeV}$ is the electron mass.
[If one restore all units a factor $\hbar^3/c^5$ appears in
(\ref{e7}).]

The domain of validy of the above expression for $\cal L$ is that
of small wave numbers and small electromagnetic field strength, i.
e.,
\cite{novello98a,gMm}
\begin{equation}
\label{e8}
k \ll m_e\;,
\end{equation}
\begin{equation}
\label{e9}
E \ll E_{cr}\equiv \frac{m_e^2}{e}=\frac{m_e^2}{\sqrt{4\,\pi\,\alpha}}\;,
\end{equation}
\begin{equation}
\label{e10}
B \ll B_{cr}\equiv \frac{m_e^2}{e}=\frac{m_e^2}{\sqrt{4\,\pi\,\alpha}}
=8.6 \times 10^{-7}\; {\rm Tesla} = 0.0086\; {\rm Gauss}\;.
\end{equation}
[Restoring all units, a factor of $c/\hbar$  appears multipling
the right hand side of (\ref{e8}) and a factor of $c^3/\hbar$
multiplies the right hand side of equations (\ref{e9}) and
(\ref{e10}).]

The corresponding modified energy-momentum tensor becomes
\cite{novello98a}
\begin{equation}
\label{e11}
T_{\mu\nu}=-4\frac{\partial\protect{\cal{L}}}{\partial
F}{F_{\mu}}^{\alpha}F_{\alpha\nu}+\left
(\frac{\partial\protect{\cal{L}}}{\partial
F^{*}}F^{*}-\protect{\cal{L}}\right )\,g_{\mu\nu}\;.
\end{equation}

We will aply the above equations for a homogeneous and isotropic universe with line element
\begin{equation}
\label{e12}
ds^2 = dt^2-a^2(t)\left
[\frac{dr^2}{1-Kr^2}+r^2d\theta^2+r^2\sin^2\theta
\,d\phi^2\right]\;,
\end{equation}
where $K=0, \pm 1$.

Such a geometry may be generated by  electromagentic fields only if these are
considered in its average properties
\cite{novello98a,stephani,tolman30}. Using the standard spatial average
process we set \cite{novello98a}
\begin{eqnarray}
\label{e13}
<E_i> &=& 0\;,\\
\label{e14}
<B_i>&=& 0\;,\\
\label{e15}
<E_iE_j>& =& -\frac{1}{3} E^2 g_{ij}\;,\\
\label{e16}
<B_i B_j>&=&-\frac{1}{3} B^2 g_{ij}\;,\\
\label{e17}
<E_iB_j>&=& 0\;.
\end{eqnarray}
Equations (\ref{e13}) - (\ref{e17}) then imply \cite{novello98a}
\begin{equation}
\label{e18}
<F_{\mu\alpha}{F^{\alpha}}_{\nu}> =
\frac{2}{3}(E^2+B^2)U_{\mu}U_{\nu}
+\frac{1}{3}(E^2-2B^2)g_{\mu\nu}\;,
\end{equation}
where
\begin{equation}
\label{e19}
U_{\mu}=\frac{dx_{\mu}}{ds}\;.
\end{equation}

Hence, for the classical Lagrangean (\ref{e3}) the average value of the
energy-momentum tensor reduces to the form of a perfect fluid
\begin{equation}
\label{e20}
<T_{\mu\nu}> = (\rho+p)\,U_{\mu}U_{\nu} -p\,
g_{\mu\nu}\;.
\end{equation}

where the density $\rho$ and pressure $p$ have the well known form
\begin{eqnarray}
\label{e21}
\rho&=& \frac{1}{2}(E^2 +B^2)\;,\\
\label{e22}
p&=& \frac{1}{3}\; \rho\;.
\end{eqnarray}

In order to analyse the modifications implied by the use of the modified Lagrangean (\ref{e5}),
we assume that the
dominant material content is a primordial plasma in
which only the average value of the squared magnetic field $B^2$
survives, i. e., we use Eqs. (\ref{e13}) - (\ref{e17}) with $E^2=0$ \cite{novello98a}.
Then Equation (\ref{e20}) still holds, but the energy density and
pressure are now given by
\begin{eqnarray}
\label{e23}
\rho & =& \frac{1}{2}B^2\,(1-2\mu B^2)\;,\\
\label{e24}
p&=&\frac{1}{6} B^2\,(1-10\mu B^2) = \frac{1}{3}\rho-\frac{4}{3}\mu
B^4\;.
\end{eqnarray}

>From Eqs. (\ref{e7}), (\ref{e23}) and (\ref{e24}) we see that the weak energy condition
 $\rho > 0$ is obeyed if
\begin{equation}
\label{e25}
B < \frac{1}{2\sqrt{\mu}}=6.0 \times 10^{-5}\; {\rm Tesla} = 0.60\; {\rm
Gauss}\;,
\end{equation}
whereas the pressure will reach negative values only if
\begin{equation}
\label{e26}
B > 2.7\times 10^{-5}\; {\rm Tesla} = 0.27\; {\rm Gauss}\;.
\end{equation}
Note, however, that expression (\ref{e5}) will hold only if the
condition (\ref{e10}) is satisfied.
Therefore, the conditions (\ref{e10}) and (\ref{e26}) are not
compatible. We shall return to this point later.

The Einstein equations for the metric (\ref{e12}) read
\begin{equation}
\label{e28}
\frac{\ddot{a}}{a}=\frac{\Lambda(t)}{3} -
\frac{4\,\pi\,G}{3}\,(\rho+3p)\;,
\end{equation}
\begin{equation}
\label{e29}
\frac{\dot{a}^2}{a^2}
+\frac{K}{a^2}=\frac{8\,\pi\,G}{3}\,\rho+\frac{\Lambda(t)}{3}\;,
\end{equation}
from which  the energy conservation equation can be written as
\begin{equation}
\label{e30}
\dot{\rho} +3\,\frac{\dot{a}}{a}\,(\rho
+p)=-\frac{\dot{\Lambda}}{8\,\pi\,G}\;.
\end{equation}
In the above equations we have allowed $\Lambda$ to be time-dependent
and the overdot means derivative with respect to the cosmic time $t$.

Recent observations that have led to the so called age problem
[13 -- 15]
have driven attention to
cosmological models with a non-vanishing $\Lambda$. Estimates of the density parameter
and some kinematical tests also point to the existence of an
effective vacuum component \cite{lima96,krauss98}. Even more
recently,
some evidence for an accelerated  cosmic expansion has arised from
measurements involving type 1a supernova at high redshifts
\cite{perlmutter98,kirshner98}.
On the other
hand, quantum field theory and inflationary cosmologies have
pointed out to the possibility of treating $\Lambda$ is a dynamical
quantity [4,16,20 -- 22]
Several mechanisms
have been identified as possible sources of fluctuating vacuum
energy (see, for example, \cite{overduin98} and references
therein).

As it will be shown below, the field equations lead
to a possible solution corresponding to a constant magnetic field
plus a time-dependent cosmological term. In this case the time
dependence of $\Lambda$ is univocally determined.

Replacing (\ref{e23}) and (\ref{e24}) in the Einstein equations
(\ref{e28}) - (\ref{e29}) and in the energy conservation equation
(\ref{e30}), we get
\begin{equation}
\label{e31}
\frac{\ddot{a}}{a}=\frac{\Lambda}{3} -
\frac{4\,\pi\,G}{3}\,B^2\,(1-6\,\mu\,B^2)\;,
\end{equation}
\begin{equation}
\label{e32}
\frac{\dot{a}^2}{a^2}
+\frac{K}{a^2}=\frac{4\,\pi\,G}{3}\,B^2\,(1-2\,\mu\,B^2)+\frac{\Lambda}{3}\;,
\end{equation}
\begin{equation}
\label{e33}
B\,\left (1-4\,\mu\,B^2\right )\,\left
(\dot{B}+2\,\frac{\dot{a}}{a}\,B\right)=-\frac{\dot{\Lambda}}{8\,\pi\,G}\;.
\end{equation}

By solving any two of the above equations we find the complete
cosmological solution for our model.
\section{GENERAL SOLUTION FOR ${\bf \Lambda = 0}$ AND A TIME-DEPENDENT MAGNETIC FIELD}
\label{s3}
This is the case studied in \cite{novello98a}
where a particular solution for the scale factor was found.
We will find the corresponding general solution and indicate how
to recover this previous result.

If $B$ is time-dependent and $\Lambda$ is a constant, Equation
(\ref{e33}) can be easily integrated to give
\begin{equation}
\label{e34}
B(t) = B_0 \,\left (\frac{a_0}{a}\right )^2\;,
\end{equation}
where $B_0$ is a constant of integration. In this paper, the
subscript $0$ does {\em not} indicate the present day value of a
quantity. Rather, it indicates the value of that quantity in an
arbitrary time $t_0$, which will appear in the general solution
for $a(t)$ as a second constant of integration. It is this second constant $t_0$
that was arbitrarily chosen in \cite{novello98a}. Thus $B_0 =
B(t_0)$ and $a_0=a(t_0)$. (In Ref. \cite{novello98a} the authors
normalized $a_0$, so that  $a_0=1$.)

Note that (\ref{e34}) holds for an arbitrary constant $\Lambda$.
If $\Lambda = K = 0$, Eqs. (\ref{e32}) and (\ref{e34}) allow us to
find $a(t)$. The general solution is
\begin{equation}
\label{e35}
a(t)= a_0\,\left [
4\,\alpha_0^2\,(t-t_0)^2\,+\,4\,\alpha_0\,\beta_0 \,(t-t_0)\,+\,1\right]^{1/4}\;,
\end{equation}
where we have defined
\begin{equation}
\label{e36}
\alpha_0\equiv \sqrt{\frac{4\,\pi\,G}{3}}\,B_0\;,
\end{equation}
\begin{equation}
\label{e37}
\beta_0\equiv \sqrt{1-2\,\mu\,B_0^2}\;.
\end{equation}

In order to compare with the results of \cite{novello98a} we
recast (\ref{e35}) in the form
\begin{equation}
\label{e38}
a(t)=a_0\,(4\,\alpha_0^2\,t^2\,+\,4\,\alpha_0\,\gamma_0\,t\,+\,\delta_0)^{1/4}\;,
\end{equation}
with
\begin{equation}
\label{e39}
\gamma_0\equiv\beta_0-2\alpha_0t_0\;,
\end{equation}
\begin{equation}
\label{e40}
\delta_0\equiv 4\alpha_0t_0\,(\alpha_0t_0-\beta_0)+1\;.
\end{equation}

The linear term in $t$ inside the parenthesis of (\ref{e38}) does not appear
in the solution given by the authors of Ref. \cite{novello98a}. This is due to the
fact that they have arbitrarily chosen the constant of integration
$t_0$ to be
\begin{equation}
\label{e41}
t_0 = \frac{\beta_0}{2\,\alpha_0}=\frac{1}{2\,B_0}\,\sqrt{\frac{3\,(1-2\,\mu\,B_0^2)}
{4\,\pi\,G}}\;.
\end{equation}

>From the solution above we may determine the time behaviour of
the magnetic field
\begin{equation}
\label{e42}
B(t)=\frac{B_0}{(4\,\alpha_0^2\,t^2\,+\,4\,\alpha_0\,\gamma_0\,t\,+\,\delta_0)^{1/2}}\;.
\end{equation}
The energy density and pressure are obtained through
the use of Eqs. (\ref{e23}), (\ref{e24}), respectivelly.

Note also that the Hubble parameter is
\begin{equation}
\label{e43}
H=\frac{\dot{a}}{a}=\frac{\alpha_0\,\left
[2\alpha_0(t-t_0)+\beta_0\right]}{\left[4\alpha_0^2(t-t_0)^2+4\alpha_0\beta_0(t-t_0)+1\right]}
\;.
\end{equation}
Hence,
\begin{equation}
\label{e44}
H_0\equiv H(t_0)
=\alpha_0\,\beta_0=B_0\,\sqrt{\frac{4\,\pi\,G\,(1-2\,\mu\,B_0^2)}{3}}\;.
\end{equation}
(In \cite{novello98a,novello98b}, the notation $H$ is used to
represent the magnetic field.)

>From (\ref{e38}) we see that, for large $t$, we recover the usual
solution for radiation dominated universes, $a(t)\propto t^{1/2}$.
However, the more interesting feature of (\ref{e38}) is that the
quadratic function inside the parenthesis does not have real
roots, being positive for any $t$. Therefore, the model is
non-singular with $a(t)$ reaching the minimum value
\begin{equation}
\label{e45}
a_{min} = a_0\,(2\,\mu\,B_0^2)^{1/4}
\end{equation}
at
\begin{equation}
\label{e46}
t_{min}=-\frac{\gamma_0}{2\alpha_0}=t_0-\frac{\beta_0}{2\alpha_0}\;.
\end{equation}

The universe thus obtained is a bouncing one: it begins
arbitrarily large at $t\ll t_{min}$, decreases until the minimum
value (\ref{e45}) at $t_{min}$ and then begins to expand.

A crucial feature of the above results is that the magnetic field at $t_{min}$ is
\begin{equation}
\label{e47}
B(t_{min}) = \frac{1}{\sqrt{2\mu}}= 6.0\times 10^{-5} \;{\rm Tesla} = 0.60
\; {\rm Gauss}\;,
\end{equation}
a value much greater than the minimal critical value given by Eq.
(\ref{e10}) which limits the domain of validity of
the first order correction given by (\ref{e5}). We are forced to conclude that the
above solution does not apply all the way back to the
point of minimum size of the universe.
This point has not been
acknowledged explicitly by the authors of reference \cite{novello98a},
although their numerical calculations \cite{novello98b,novellopc} based on the exact effective
Lagrangean density (beyond the first order approximation) seems to
confirm the non-singular behaviour of the model. We should remark
that, in \cite{novello98a}, the requirement (\ref{e8}) is
mentioned, but not the condition (\ref{e10}), which is established
only in equation (6.84) of \cite{gMm}.
Note also that
\begin{equation}
\label{e27}
\rho(t_{min}) = 0\;.
\end{equation}

The times for which the constraint (\ref{e10}) is obeyed may be
easily determined using (\ref{e42}). We find that the solution
(\ref{e38}) fulfill  the above requirement for
\begin{equation}
\label{e48}
t \, < \, t_{min} - t_{(1)}\hspace{1cm} {\rm and}\hspace{1cm} t\,
> t_{min} + t_{(1)}\;,
\end{equation}
where
\begin{equation}
\label{49}
t_{(1)} =
m_{pl}\,\sqrt{\frac{3}{16\,\pi}\left(\frac{1}{B_{cr}^2}-2\,\mu\right
)}\approx\frac{m_{pl}}{4B_{cr}}\sqrt{\frac{3}{\pi}}=3.4\times10^{24}\;{\rm
GeV}^{-1} = 2.2\;{\rm s}\;,
\end{equation}
where $m_{pl}= G^{-1/2}=1.2211\times 10^{19}\,{\rm GeV}$ is the
Planck mass.

This seems to be a disappointing conclusion about the capability
of the first order quantum correction (\ref{e5}) to describe the
very early stages of the universe. In the next two sections we
try  to evade this problem by allowing for the existence of a non-vanishing
cosmological term, as it has been increasingly indicated by recent
theoretical and observational results \cite{krauss98}.

One could speculate that models with a time varying fine structure
``constant'' could perhaps evade the above mentioned problem. In fact,
if $\alpha$, and consequently $\mu$, was
grater at early times, the value of $B(t_{min})$ could reach values
below the constraint (\ref{e10}). However, in spite of the fact that
models with
oscillatory variation of the fundamental constants have been proposed
\cite{barrow87,damour94},
recent calculations
using the Keck Telescope data seems to indicate a negative variation of  $\alpha$
at high redshift ($z> 1$): $\Delta\,\alpha /\alpha = -1.1\pm
0.4\times 10^{-5}$ \cite{barrow98a,barrow98b}.

Before going to this next step, let us mention that, if (\ref{e38})
could describe the entire evolution of the universe in the distant
past, it would imply the existence of an inflationary era ($\ddot{a} > 0$) in the interval
\begin{equation}
\label{e50}
t_{min}-t_I\, <\, t\, <\, t_{min} + t_I\;,
\end{equation}
where
\begin{equation}
\label{e51}
t_I=\frac{m_{pl}}{2}\,\sqrt{\frac{3\,\mu}{\pi}}\approx 7.1\times
10^{22}\,{\rm GeV}^{-1}=0.046\,{\rm s}\;.
\end{equation}

Figure 1 shows the scale factor, the magnetic field, the energy
density and the pressure as a function of time for a definite value
of $B_0$. The time interval where the weak field approximation
breaks down is also indicated.
\section{GENERAL SOLUTION FOR A CONSTANT NON-VANISHING $\bf
\Lambda$ AND A TIME-DEPENDENT MAGNETIC FIELD}
\label{s4}
We will now extend the analysis made in \cite{novello98a}
and investigate how the above conclusions are modified with the
introduction of a constant cosmological term.

For $K=0$ and $\Lambda = {\rm constant}\neq 0$, substitution of (\ref{e34}) into
(\ref{e32}) leads to the equation
\begin{equation}
\label{e52}
\dot{Z}^2=16\,\left[\lambda Z^2+\alpha_0^2\,(Z-2\mu B_0^2)\right ]\;,
\end{equation}
where we have defined
\begin{equation}
\label{e53}
Z\equiv\left (\frac{a}{a_0}\right)^4\;,
\end{equation}
\begin{equation}
\label{e54}
\lambda\equiv\frac{\Lambda}{3}\;.
\end{equation}
Equation (\ref{e52}) can be easily integrated to give
\begin{equation}
\label{e55}
a(t) = a_0 \left(\frac{1}{4\lambda}\right )^{1/4}\,\left
[C_0\,e^{4\sqrt{\lambda}\,(t-t_0)}+\frac{D_0}{C_0}\,e^{-4\sqrt{\lambda}\,(t-t_0)}-2\,
\alpha_0^2\right ]^{1/4}\;,
\end{equation}
where
\begin{equation}
\label{e56}
C_0\equiv
\alpha_0^2+2\lambda+2\sqrt{\lambda\,(\lambda+\alpha_0^2-2\,\alpha_0^2\,\mu\,B_0^2)}\;,
\end{equation}
\begin{equation}
\label{e57}
D_0\equiv \alpha_0^2\,(\alpha_0^2+8\,\lambda\,\mu\,B_0^2)\;.
\end{equation}

The Hubble parameter is
\begin{equation}
\label{e58}
H(t) = 2B_0\sqrt{\lambda}\,\left [C_0\, e^{4\sqrt{\lambda}\,(t-t_0)}
+\frac{D_0}{C_0}\,
e^{-4\sqrt{\lambda}\,(t-t_0)}-2\,\alpha_0^2\right ]^{-1/2}\;.
\end{equation}

It is straightforward to see that the term inside the square
brackets of (\ref{e55}) is positive for all $t$ and that the scale
factor reaches its minimum value
\begin{equation}
\label{e59}
a_{min} = a_0\,\left [\frac{\alpha_0}{2\,\lambda}\,\left (
\sqrt{\alpha_0^2+8\,\lambda\,\mu\,B_0^2}-\alpha_0\right )\right
]^{1/4}
\end{equation}
at
\begin{equation}
\label{e60}
t_{min}= t_0 + \frac{1}{8\sqrt{\lambda}}\,\ln\left
(\frac{D_0}{C_0^2}\right )\;.
\end{equation}

As in the previous case, the universe bounces at $t_{min}$ and, if
the solution would effectivelly hold near $t_{min}$, an
inflationary phase would take place for all values of $t$ such
that
\begin{equation}
\label{e61}
C_0^2 x^4-8\alpha_0^2C_0x^3+14D_0x^2-8\alpha_0^2\frac{D_0}{C_0}x
+\frac{D_0^2}{C_0^2} \, >\, 0\;,
\end{equation}
where
\begin{equation}
\label{e62}
x\equiv e^{4\sqrt{\lambda}\,(t-t_0)}\;.
\end{equation}

Nevertheless, the magnetic field at $t_{min}$ is
\begin{equation}
\label{e63}
B(t_{min}) = \left [\frac{\Lambda}{2\,\pi\,G\,\left
(\sqrt{1+\frac{2\,\Lambda\,\mu}{\pi\,G}}-1\right)}\right
]^{1/2}\;.
\end{equation}

>From the above expression we see that $B(t_{min})\rightarrow
\frac{1}{\sqrt{2\mu}}$ as $\Lambda \rightarrow 0$ and that
$B(t_{min})\rightarrow \infty$ as $\Lambda\rightarrow\infty$.
Therefore this model suffers from the same problem as the previous
one: the first order approxiamtion (\ref{e5}) can not describe the
dynamics all the way back to the point of minimum compression.

>From the condition $B(t) < B_{cr}$, we find that the solution is
valid for any time $t$ such that
\begin{equation}
\label{e64}
t-t_{min} =
\frac{1}{4}\,\sqrt{\frac{3}{\Lambda}}\ln\left[\frac{A}{B_{cr}^2\,\sqrt{4\,\pi\,G\,(\pi\,G+2\,
\mu\,\Lambda)}}\right ]\;,
\end{equation}
where
\begin{equation}
\label{e65}
A\equiv \Lambda +2\,\pi\,G\,B_{cr}^2\pm\sqrt{\Lambda\,\left
[\Lambda+4\,\pi\,G\,B_{cr}^2\,(1-2\,\mu\,B_{cr}^2)\right ]}\;.
\end{equation}
The domain of validity of the solution will then depend on the
value of $\Lambda$.

Figure 2 and Figure 3 show the scale factor, the magnetic field, the energy
density and the pressure as a function of time, for some values of
$\Lambda$ and $B_0$.
\section{GENERAL SOLUTION FOR A CONSTANT MAGNETIC FIELD AND A TIME-DEPENDENT
$\bf \Lambda$}
\label{s5}
We will now turn to the case when the magnetic field does not vary with time.
This solution of Eq. (\ref{e33}) has not been analysed  in
Ref. \cite{novello98a}. If $\dot{\Lambda} = 0$, this leads to
the well known models $p=\rho=0$ or $p=-\rho = {\rm constant}$.
However, if we treat $\Lambda$ is a dynamical variable, then (\ref{e33})
with
\begin{equation}
\label{e66}
B(t)= B_0 = {\rm constant}\;,
\end{equation}
leads to
\begin{equation}
\label{e67}
\dot{\Lambda}=3K_0\frac{\dot{a}}{a}
\end{equation}
and hence
\begin{equation}
\label{e68}
\Lambda(t)=\Lambda_0 + 3 K_0\ln\left(\frac{a}{a_0}\right )\;,
\end{equation}
where
\begin{equation}
\label{e69}
\Lambda_0 \equiv \Lambda (t_0)
\end{equation}
and
\begin{equation}
\label{e70}
K_0 \equiv -\frac{16\,\pi\,G}{3}\,B_0^2\,(1-4\,\mu\,B_0^2)\;.
\end{equation}

Substituing this result into (\ref{e32}), we get for the scale factor
\begin{eqnarray}
\label{e71}
a(t)& = & a_0\, \exp\left
[\frac{K_0}{4}\,(t-t_0)^2\,+\,H_0\,(t-t_0)\right ]\nonumber\\
 & = & a_0\, \exp\left(\frac{K_0}{4}\,t^2+\beta_1\,
t+\beta_0\right )\;,
\end{eqnarray}
where
\begin{eqnarray}
\label{e72}
H_0& \equiv &
\sqrt{\frac{\Lambda_0}{3}+\frac{4\,\pi\,G}{3}\,B_0^2\,(1-2\,\mu\,B_0^2)}\nonumber\\
 & =&\sqrt{\frac{\Lambda_0}{3}+\frac{8\,\pi\,G}{3}\,\rho}\;,
\end{eqnarray}
\begin{equation}
\label{e73}
\beta_1\equiv -\left(\frac{K_0}{2}\,t_0-H_0\right )\;,
\end{equation}
and
\begin{equation}
\label{e74}
\beta_0\equiv t_0\,\left (\frac{K_0}{4}\,t_0-H_0\right )\;.
\end{equation}

The Hubble parameter is
\begin{eqnarray}
\label{e75}
H(t)&=&\frac{K_0}{2}\,(t-t_0) + H_0\nonumber\\
 &=&\frac{K_0}{2}\,t+\beta_1\;,
\end{eqnarray}
and we have $H=0$ for
\begin{equation}
\label{e76}
t_c= -\frac{2\beta_1}{K_0}=t_0-\frac{2H_0}{K_0}\;.
\end{equation}
At this point, the scale factor reaches the value
\begin{equation}
\label{e77}
a(t_c)=a_0\,e^{-H_0^2/K_0}\;,
\end{equation}
whereas the cosmological term is
\begin{equation}
\label{e78}
\Lambda(t_c)=\Lambda_0-3H_0^2=-4\,\pi\,G\,B_0^2\,(1-2\,\mu\,B_0^2)=-8\,\pi\,G\,\rho\;.
\end{equation}

The behaviour of the solution will depend on the sign of the constant $K_0$
(for $K_0 = 0$ we get the de Sitter solution).

For $K_0 >0$, that is, for $B_0 > 1/(2\sqrt{\mu})$, the universe
begins arbitrarily large as $t\rightarrow - \infty$, contracts
until reaches its minimum size at $t_c$ and begins to expands
without limit from this point on. The solution is always
acelerated ($\ddot{a}>0$). Note that this case violates condition
(\ref{e10}).

A much more interesting solution is the one corresponding to $K_0<0$
($B_0 <1/(2\sqrt{\mu})\approx 4.2\times 10^{-5}\:{\rm
Tesla}=0.42\:{\rm Gauss}$), since we can make $B_0$ satisfy
(\ref{e10}) in this case. For this range of $B_0$, $a(t)$ approaches zero
asymptotically as $t\rightarrow -\infty$, reaches a {\em maximum}
at $t_c$ and begins to contract, approaching zero as $t\rightarrow
+\infty$. Moreover, $\ddot{a} > 0$ for
\begin{equation}
\label{e79}
t< t_c-\sqrt{-\frac{2}{K_0}}\hspace{1cm}{\rm and}\hspace{1cm}
t> t_c+\sqrt{-\frac{2}{K_0}}\;.
\end{equation}

It is worthy remarking that the time interval $\Delta t_{(NI)}$,
prior to $t_c$, for which the solution is {\em not} inflationary
depends on the value of $B_0$ as
\begin{equation}
\label{e79b}
\Delta t_{(NI)} = \sqrt{-\frac{2}{K_0}}=
\frac{1}{B_0}\,\sqrt{\frac{3}{8\,\pi\,G\,(1-4\,\mu\,B_0^2)}}\;.
\end{equation}
This time interval reaches a minimum at $B_0^2=1/(8\mu)$ and goes
to infinity as $B_0\rightarrow 0$ or $B_0^2\rightarrow 1/(4\mu)$.
Remembering that, for consistency, condition (\ref{e10}) must be
obeyed, we see that this time interval may be as small as
$3.2\:{\rm s}$. Hence we have a cosmological model that could
describe a universe that inflates through much of its history.

Note that, as far as $\Lambda(t)$ is concerned, $t_c$ is a point of minimum for both
$K_0>0$ and $K_0< 0$. For the case of interest, ($K_0<0$), the
evolution proceeds in the  following way (note that the weak energy condition
($\rho >0$) is automatically guaranteed if $K_0 < 0$). In the far past,
the universe is arbitrally small,
with an arbitrarilly large vacuum energy.  It expands
in such a way that during this expansion phase, the
vacuum energy decays so as to keep
$\rho$ constant. The vacuum energy eventually becomes negative at
the time
\begin{equation}
\label{e80}
t_1=t_c+\frac{2}{K_0}\,\sqrt{H_0^2-\frac{\Lambda_0}{3}}\;,
\end{equation}
reaching its minimum value
\begin{equation}
\label{e81}
\rho_{vac}^{(min)}=\frac{\Lambda_0-3H_0^2}{8\,\pi\,G}
\end{equation}
at $t_c$, where the universe reaches its maximum size. Thereafter,
the contraction phase takes place. The vacuum energy then grows,
becoming positive at
\begin{equation}
\label{e82}
t_2=t_c-\frac{2}{K_0}\,\sqrt{H_0^2-\frac{\Lambda_0}{3}}\;,
\end{equation}
whereas $\rho$ remains constant. The expected tendency of $\rho$ growing adiabatically
with the decreasing $a$ is compensated by the increase
in $\rho_{vac}$. The universe would end up arbitrarily small as
$t\rightarrow\infty$ with an arbitrarilly large vacuum energy.

Figure 4 shows the scale factor and the cosmological
term as a function of time for $K_0 > 0$ and some values of $B_0$
and $\lambda_0$, where we have further defined
\begin{equation}
\label{e82b}
\lambda_0\equiv \frac{\Lambda_0}{3}\;.
\end{equation}
Figure 5 shows the same quantities for $K_0 < 0$.

If one expects any such model to properly describe
the evolution of the real universe, it would
be advisable to take into account other matter fields, such as
ultrarelativistic matter, scalar fields or dust. In
\cite{novello98a}, it was demonstrated, for the case $\Lambda=0$, $B=B(t)$,
that the presence of ultrarelativistic matter with an equation of state $p_{(ur)}
=\rho_{(ur)}/3$ would just amount for a reparametrization of the constants
$B_0$ and $\mu$. The effects of other matter fields in the models
analysed in this section are currently under investigation.
\section{SUMMARY AND CONCLUSIONS}
\label{s6}
We have examined some consequences of considering first order quantum
corrections to Maxwell elctrodynamics in zero curvature FRW
universes. We have derived general analytical solutions under
three different assumptions. When the cosmological term $\Lambda$
is identically zero and the dynamics is driven by a time dependent magnetic field,
we have obtained the general form of the
solution found previously by Novello {\em et al.} \cite{novello98a}.
We have found a new solution when $\Lambda$ is a non-vanishing
constant. In both cases, the universe is non-singular, bouncing
at a critical time when it reaches its minimum size. However, we
have shown that, near this critical time, the magnetic field
increases beyond the value allowed by the use of the first order
approximation to the effetcive Lagrangean density. The time range where
this weak field approximation does apply was evaluated. On the other hand, numerical
work, that takes into account higher order terms, indicates that the
non-singular behaviour is preserved \cite{novello98b,novellopc}.

We have derived a third solution that describes a universe driven
by a time dependent $\Lambda$ that sustains a constant magnetic
field. In this case, the time behaviour of the cosmological term
is univocally determined and depends on the logarithym of the
scale factor. For small enough values of the magnetic field
strength (so that the first order corrections can be used safely),
the universe begins arbitrarily small as $t\rightarrow\-\infty$,
expands to a maximum size $a_{max}$ at $t_c$ and then contracts back to zero
size as  $t\rightarrow\+\infty$. The energy density associated
with $\Lambda$ is arbitrarily large at  $t\rightarrow -\infty$,
reaches a minimum at $t_c$ and grows without limit as
$t\rightarrow\+\infty$. It therefore decays during the expansion
phase and increases during the contraction era, so as to keep the magnetic energy
constant. The dynamics is inflationary during most of the cosmic
history, except near $t_c$.

We are presently attempting to generalize our results for
universes with non-zero curvature as well as analysing how
our models would be modified by the presence of other matter
fields. Another possibility is to study how the time evolution for
the magnetic field, to be derived from (\ref{e33}), would be modified
by imposing a definite decaying law for $\Lambda$. This could be  chosen among
the several forms presented in the literature (see, for example,
\cite{overduin98}). More accurate results could possibly be found
by considering higher order terms in the effective Lagrangean.

\pagebreak
\begin{figure}
\vspace{.2in}
\centerline{\psfig{figure=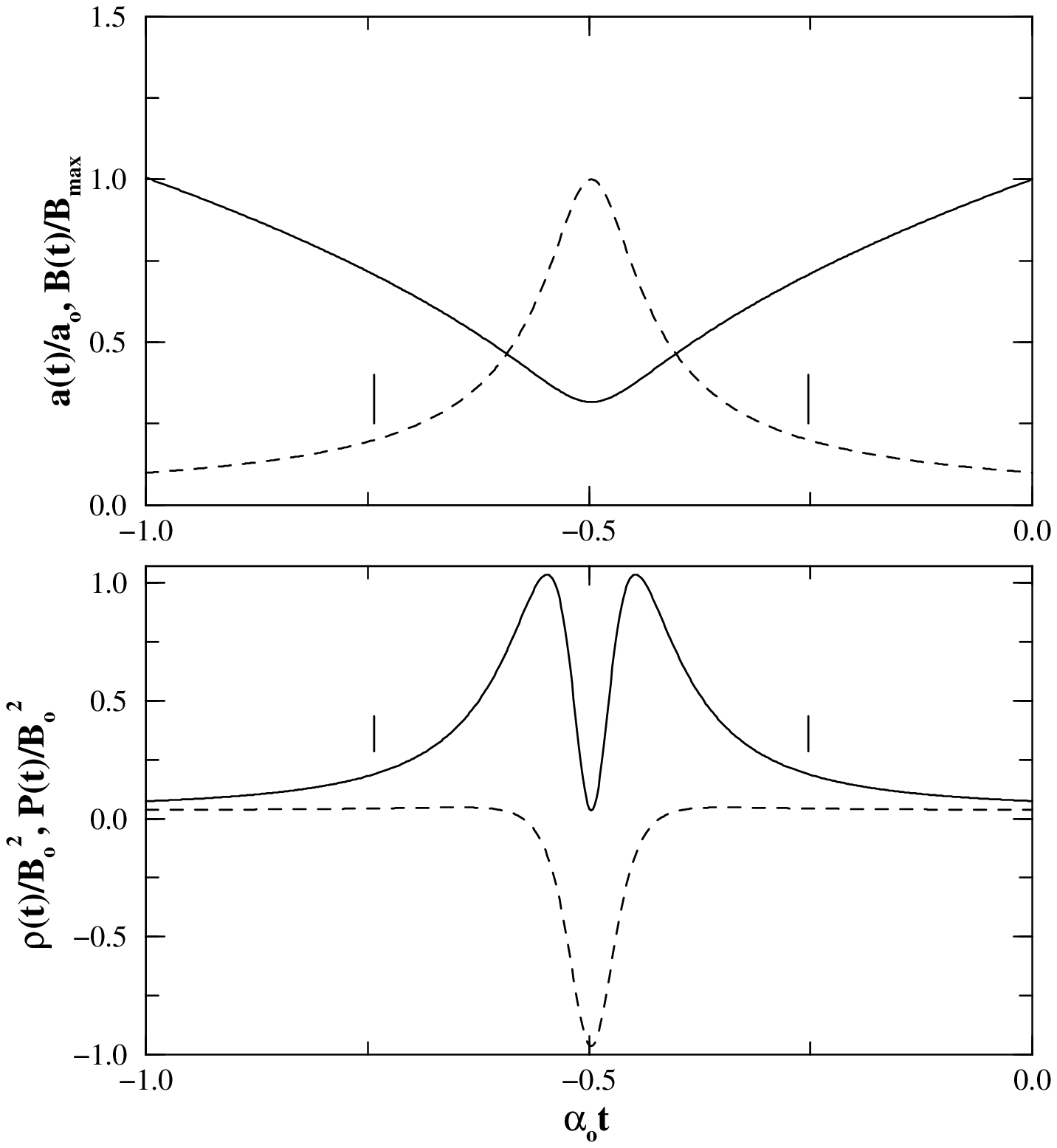,width=3truein,height=3truein}
\hskip 0.1in}
\caption{\protect{The upper panel shows the scale factor (solid line)
and the magnetic field (dashed line) while the lower panel shows
the energy density (solid line) and the pressure (dashed line) for
the model with $\Lambda =0$. The vertical bars indicate the time
interval $t_{min}-t_{(1)} < t < t_{min}+t_{(1)}$, during which the
constraint (10) is not obeyed. $B_0$ has been chosen such that
$\sqrt{2\mu}B_0 =0.2$ and $B_0/B_c=0.5$.}}
\end{figure}
\begin{figure}
\vspace{.2in}
\centerline{\psfig{figure=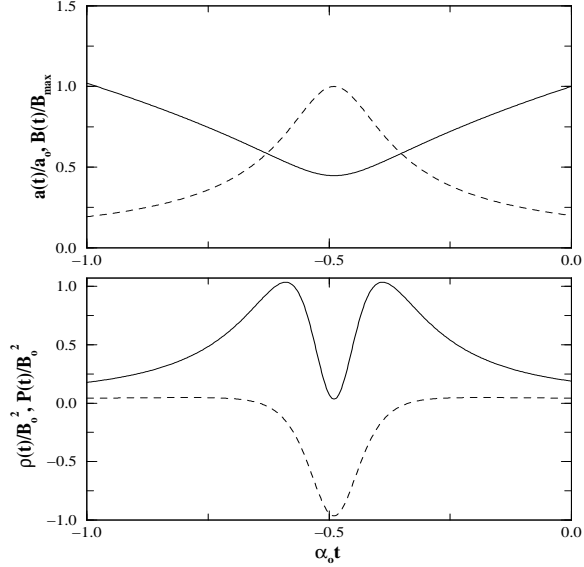,width=3truein,height=3truein}
\hskip 0.1in}
\caption{\protect{The upper panel shows the scale factor (solid line)
and the magnetic field (dashed line) while the lower panel shows
the energy density (solid line) and the pressure (dashed line) for
the model with a constant non-vanishing $\Lambda$. The values for
$\Lambda$ and $B_0$ are such that $\sqrt{\lambda}/\alpha_0 =
2\times 10^{-4}$ and $\sqrt{2\mu}B_0 =0.2$.}}
\end{figure}
\begin{figure}
\vspace{.2in}
\centerline{\psfig{figure=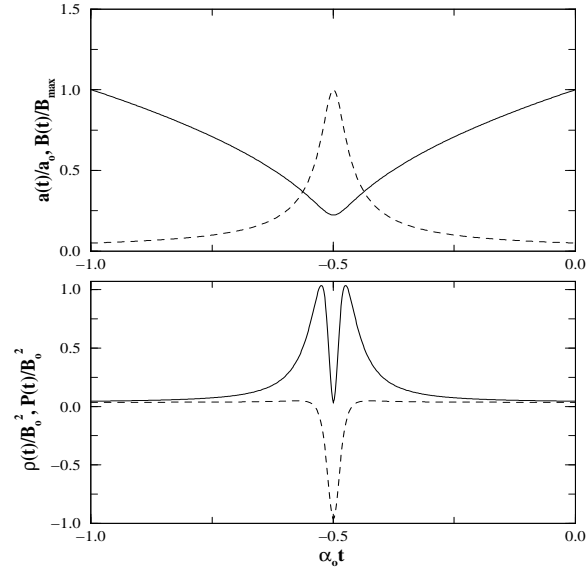,width=3truein,height=3truein}
\hskip 0.1in}
\caption{\protect{As in Figire 2 but for $\sqrt{\lambda}/\alpha_0=5\times 10^{-5}$
and $\sqrt{2\mu}B_0=0.05$.}}
\end{figure}
\begin{figure}
\vspace{.2in}
\centerline{\psfig{figure=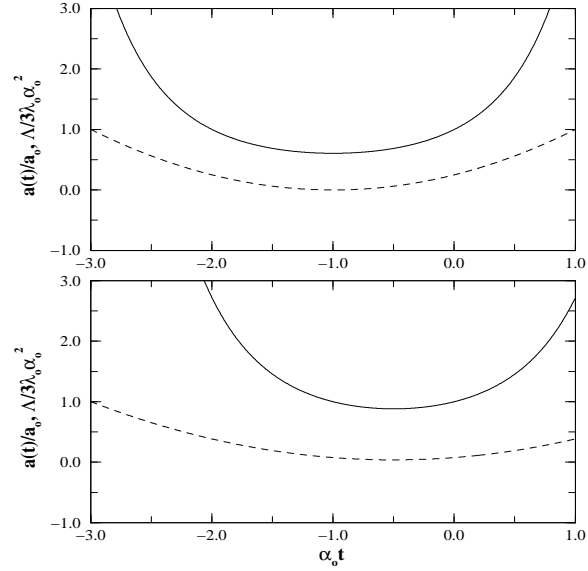,width=3truein,height=3truein}
\hskip 0.1in}
\caption{\protect{The scale factor (solid line) and the cosmological term (dashed line)
for the model with constant magnetic field, time-dependent
$\Lambda$ and $K_0 > 0$ ($\sqrt{2\mu}B_0 =1$). In the upper
panel $\sqrt{\lambda_0}/\alpha_0 =1$ and the lower panel is for
$\sqrt{\lambda_0}/\alpha_0 =0.5$.}}
\end{figure}
\begin{figure}
\vspace{.2in}
\centerline{\psfig{figure=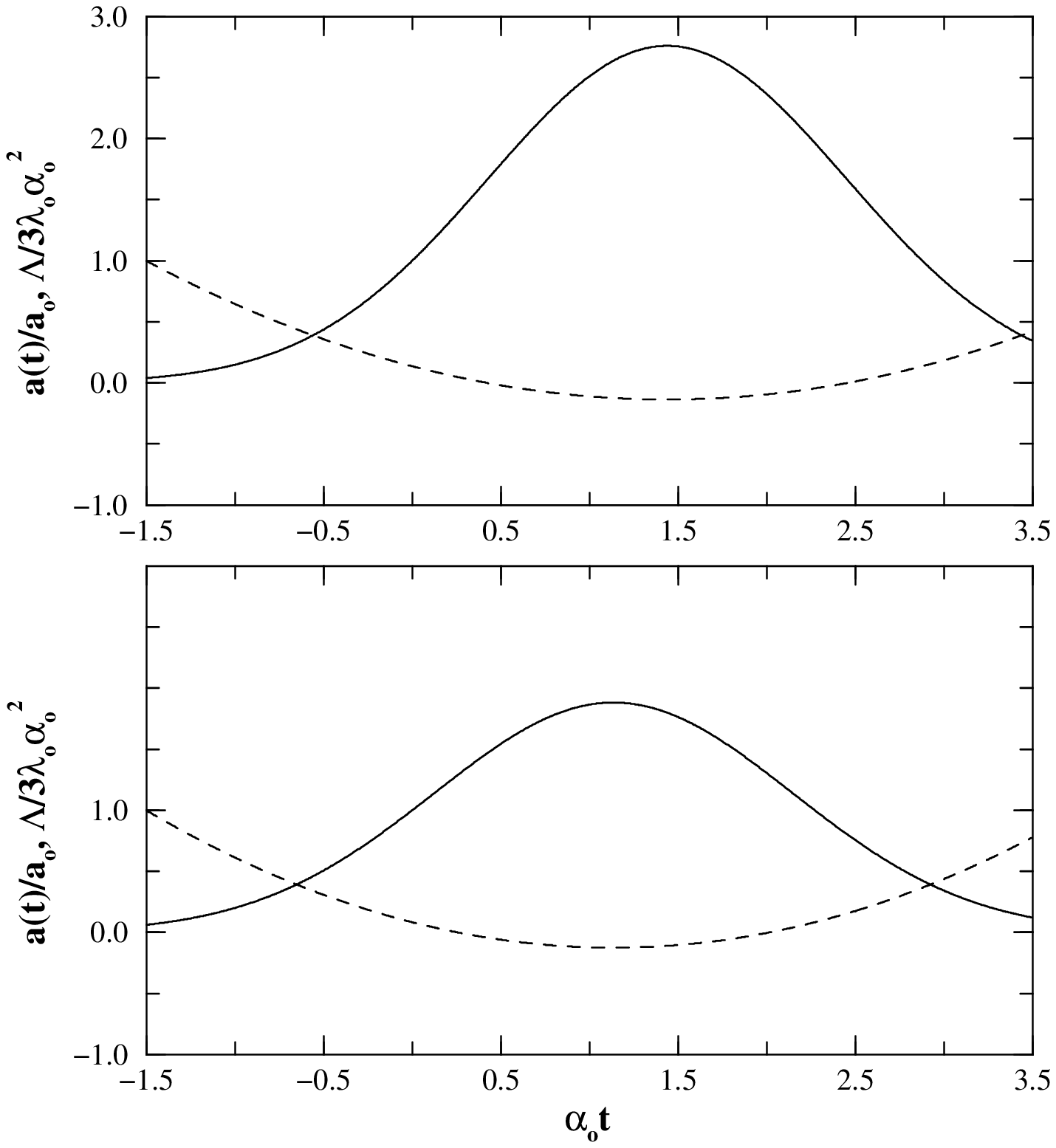,width=3truein,height=3truein}
\hskip 0.1in}
\caption{\protect{As in Figure 4 but for $K_0 < 0$ ($\sqrt{2\mu}B_0
=0.1$). In the upper
panel $\sqrt{\lambda_0}/\alpha_0 =1$ and the lower panel is for
$\sqrt{\lambda_0}/\alpha_0 =0.5$.}}
\end{figure}

\begin{references}
%
\bibitem{novello98a} M. Novello, J. M. Salim, V. A. De Lorenci,
and R. Klippert, gr-qc/9806076 (1998).
\bibitem{heisenberg36} W. Heisenberg and H. Euler, Z. Phys {\bf
98}, 714 (1936).
\bibitem{novello98b} M. Novello, J. M. Salim, V. A. De Lorenci,
and R. Klippert, gr-qc/9809080 (1998).
\bibitem{overduin98} J. M. Overduin and F. I. Cooperstock,
astro-ph/9805260 (1998).
\bibitem{suh98} I. Suh and G. J. Mathews, astro-ph/9812185 (1998).
\bibitem{olinto98} A. V. Olinto, astro-ph/9807051 (1998).
\bibitem{kT} E. W. Kolb and M. S. Turner, {\em The Early
Universe}, Addison-Wesley, Reading, 1994.
\bibitem{jackson} J. D. Jackson, {\em Classical Electrodynamics},
John Wiley, New York, 1975.
\bibitem{schwinger51} J. Schwinger, Phys. Rev. {\bf 82}, 664
(1951).
\bibitem{gMm} A. A. Grib, S. G. Mamayev, and V. M. Mostepanenko,
{\em Vacuum Quantum Effects in Strong Fields}, Friedmann
Laboratory Publishing, St. Petersburg, 1994.
\bibitem{stephani} H. Stephani, {\em General Relativity},
Cambridge University Press, Cambridge, England, 1990.
\bibitem{tolman30} R. C. Tolman and P. Ehrenfest, Phys. Rev. {\bf
36}, 1791 (1930).
\bibitem{age1} W. L. Freedman et al., Nature (London) {\bf 371}, 27 (1994).
%
\bibitem{age2} M. J. Pierce et al.,
Nature (London) {\bf 371}, 29 (1994).
\bibitem{turner97} M. S. Turner, astro-ph/9703161 (1997).
%
\bibitem{lima96} J. A. S. Lima and M. Trodden, Phys. Rev. D {\bf
53}, 4280 (1996).
\bibitem{krauss98} L. M. Krauss, hep-ph/9807376 (1998).
\bibitem{perlmutter98} S. Permutter {\em et al.}, astro-ph/9812133
(1998).
\bibitem{kirshner98} R. Kirshner {\em et al.}, astro-ph/9805201
(1998).
\bibitem{peebles88} P. J. Peebles and B. Ratra, Ap. J. {\bf 325},
L17 (1988).
\bibitem{chen90} W. Chen and Y. Wu, Phys. Rev. D {\bf 41}, 695
(1990).
\bibitem{carvalho92} J. C. Carvalho, J. A. S. Lima, and I. Waga,
Phys. Rev. D {\bf 46}, 2404 (1992).
\bibitem{novellopc} M. Novello, private communication (1999).
\bibitem{barrow87} J. D. Barrow, Phys. Rev. D {\bf 35}, 1805
(1987).
\bibitem{damour94} T. Damour and A. M. Polyakov, Nucl. Phys. B {\bf 423},
532 (1994).
\bibitem{barrow98a} J. K. Webb, V. V. Flambaum, C. W. Churchill,
M. J. Drinkwater, and J. D. Barrow, astro-ph/9803165 (1998).
\bibitem{barrow98b} J. D. Barrow and J. Magueijo, astro-ph/9811072
(1998).
\end{references}
\end{document}